# Automatic Financial Trading Agent for Low-risk Portfolio Management using Deep Reinforcement Learning

Wonsup Shin, Seok-Jun Bu and Sung-Bae Cho*

Department of Computer Science, Yonsei University, Seoul, South Korea

{wsshin2013, sjbuhan, sbcho}@yonsei.ac.kr

**Abstract**

The autonomous trading agent is one of the most actively studied areas of artificial intelligence to solve the capital market portfolio management problem. The two primary goals of the portfolio management problem are maximizing profit and restrainting risk. However, most approaches to this problem solely take account of maximizing returns. Therefore, this paper proposes a deep reinforcement learning based trading agent that can manage the portfolio considering not only profit maximization but also risk restraint. We also propose a new target policy to allow the trading agent to learn to prefer low-risk actions. The new target policy can be reflected in the update by adjusting the greediness for the optimal action through the hyper parameter. The proposed trading agent verifies the performance through the data of the cryptocurrency market. The Cryptocurrency market is the best test-ground for testing our trading agents because of the huge amount of data accumulated every minute and the market volatility is extremely large. As a experimental result, during the test period, our agents achieved a return of 1800% and provided the least risky investment strategy among the existing methods. And, another experiment shows that the agent can maintain robust generalized performance even if market volatility is large or training period is short.

## 1. Introduction

The autonomous trading agent is one of the most actively discussed fields in decades prior to modern artificial intelligence. This field is attracting attention in many other areas, including online auction [1], [2], and energy market [3]. Especially, it has been extensively studied in the financial market [4], [5] for portfolio management problem.

An autonomous trading agent for portfolio management problem has two primary goals. The first is to raise high profits and the second is to restraint risk [6]. Although the ultimate goal of an investment is to raise high profits, many books and studies emphasize the importance of the second primary goal. For example, first, some empirical results show that the positive correlation between risk and profit is very weak [7], [8], and even pursuing low risk can yield higher returns [9]. Second, the study of [10] shows that when the risk of investment increases, customer satisfaction decreases. In addition, global investment professionals emphasize minimizing the risk of investment through careful analysis through their books or quote [11], [12]. These studies and references appeal the need for low risk portfolio management agent research.

There have been many machine learning works to develop an autonomous trading agent, or at least to predict capital market volatility [13]. The mainstream of the works was the ones using supervised learning(SL) algorithms [14], [15]. The SL-based methods mainly learn an agent through a statistical model designed by human knowledge. However, compared to the conventional learning tasks, the dynamics of the capital market were more challenging to express as the domain knowledge of the human [16]. As another stream of research, there are studies using variants of Reinforcement Learning(RL) [17], [18]. The RL-based methods have the advantage that domain knowledge is not required and only state, action, and reward specifications are required. In addition, successful deep RL-based studies that solved portfolio management problem using the Deep Q-Network(DQN) method [19], which has recently achieved tremendous success in video games, further highlight the importance of RL-based research [20], [21]. But both SL-based and RL-based methods focused solely on maximizing profit rather than risk restraint.

In this paper, we propose new deep RL-based method for portfolio management problems. The main contribution of this paper is to propose a trading agent that can manage the portfolio in the capital market considering both profit maximization and risk restraint. The proposed agent can manage multiple assets together and the number of assets is scalable. The assets to be used are preselected by the user and the information of selected assets are preprocessed to form a state. Also, the ratio to which the selected action is applied is continuously selected according to the softmax value for the q-value. The target policy of the proposed deep RL agent allows the user to arbitrarily adjust the greediness for the action with the highest q-value. By appropriately reducing the greediness of the target policy, our agent takes the risk into account of exploration.

We verify the validity of the proposed method through the beck-test in the cryptocurrency market. The cryptocurrency market began in 2008 with the issuance of bitcoin, the electronic and decentralized alternatives to government-issued moneys [22]. The cryptocurrency market is the best test-ground for testing our method. Because, first, the cryptocurrency market is extremely volatile. Unlike the stock market, cryptocurrency is easy to buy through the online exchange 24 hours a day, and there are no bounds of maximum price fluctuation per day. Therefore, the cryptocurrency market has an enormous risk both empirically and potentially. Second, the cryptocurrency market has enormous and diverse data. As shown in Fig 1, the cryptocurrency market has grown explosively, with tremendous attention. For example, if you bought $ 1 bitcoin on October 27, 2010, you can sell it for $ 103,453 on January 31, 2018 [23]. Thanks to this explosive interest, more than 1500 altcoins, which represent all virtual currency issued since bitcoin, have been issued and new data for every coin are being stored every minute.

As the result of our experiments, our trading method achieved a profit of 1877% during the test period and achieved the most stable profit among high income earning methods. Surprisingly, we also had a profit of about 115% in experiments on the period when the market price fell significantly to -49%. In order to verify the robustness of our method, agents were learned for different periods and tested for the same period, respectively. Although the portfolio management ability became unstable as the learning period became shorter, the agent gained positive returns in all cases.

Section 2 discusses previous studies on capital market forecasting and autonomous trading agents. In particular, we will focus on studies that set the cryptocurrency market as a domain. Section 3 describes the overall architecture, problem definition, preprocessing method, target policy, and network structure of the proposed RL-based autonomous trading agent. Section 4 uses cryptocurrency data to compare the performance of the proposed method with other methods. In addition, the significance of the proposed method is verified through various experiments. Section 5 summarizes the overall content and describes further studies.

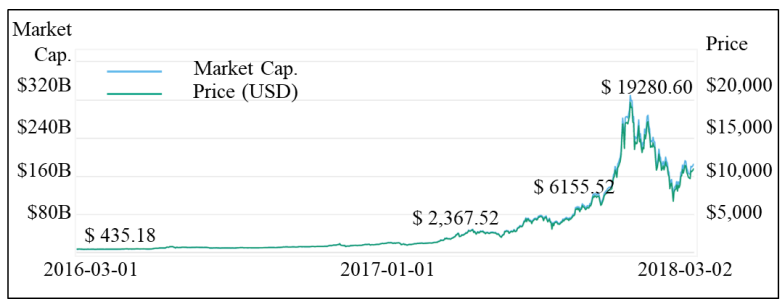

**Fig. 1. The volatility of the bitcoin values in resent two years**

## 2. Related Works

In this section, we introduce various related works based on machine learning approaches for comparison with the proposed trading agent. In addition, several studies are included to introduce various frameworks for forecasting capital market volatility.

Jae Won Lee predicted the stock price fluctuation by applying the temporal difference (TD) algorithm [24]. This was a pioneering study applying the RL-based method using the q-function approximated by the neural network(NN) to the capital market prediction. Huang et al. estimated the weekly stock price volatility using a Support Vector Machine(SVM) algorithm [25]. The study achieved 75% accuracy and demonstrated the feasibility of a trading agent. Schumaker et al. estimated stock price volatility using external information related to capital markets [26]. They analyzed text data existing on the web and used it as part of the features of the SL-based method. Simulation trading experiments also showed that the trading agent that applied the method could make a profit. Bin Li et al. proposed a framework to provide a portfolio management strategy for multi-assets, called Passive Aggressive Mean Reversion (PAMR), using the mean reversion attribute of the capital market [27]. Trading agents running in the proposed framework showed state-of-the-art performance in those days. Pater et al. applied various SL-based methods to the stock market volatility forecasting problem and compared them [28]. Despite uncertainties in stock market volatility, they showed an accuracy of 83% in experiments using Random Forest(RF) methods.

In the mid of 2010, many studies using the cryptocurrency market as a domain have been introduced. McNally et al. encoded the cryptocurrency market using a wavelet transform and estimated the variability by applying the encoded features to the Recurrent Neural Network (RNN) with LSTM [29]. They achieved about 50% of classification accuracy, but contributed to the modeling sequence of cryptocurrency market. Zbikowski et al. first applied a SL-based trading agent to the cryptocurrency market [30]. They showed that the trading agent could also work in the cryptocurrency market. Amjad et al. proposed frameworks for predicting volatility in the cryptocurrency market using statistical techniques [31]. trading agents using frameworks with statistical techniques gained more than six times over the test period. Unlike previous approaches, Jiang et al. do not include a volatility prediction framework inside the agent. Instead, they propose a trading agent using a deep reinforcement learning (deep RL) architecture [32]. Deep RL-based methods such as DQN directly map the input state into output action. As a result, deep RL-based methods can learn without the incomplete domain knowledge of human.

Most of all, the agents of all the studies presented did not take the risks into account from the investment. On the contrary, our trading agent considers both profit maximization and risk restraint. Table 1 contains a summary of the methods discussed above.

**Table 1. Related works on trading agent using machine learning algorithms**

| Year | Authors | Methods | Description |
|---|---|---|---|
| 2001 | Jae Won Lee [24] | TD(0), NN | Predict volatility of stock market with RL-based |

| Year | Authors | Methods | Description |
|------|---------|---------|-------------|
|      |         |         | method |
| 2005 | Huang [25] | SVM | Predict weakly volatility of stock market with SL-based method |
| 2009 | Schumaker [26] | SVM | Model stock volatility with external information |
| 2012 | Bin Li [27] | PAMR | Apply the mean reversion property to predict the volatility of financial market |
| 2015 | Pater [28] | NB, RF, SVM, NN | Compare stock market forecasting ability of various SL-based methods |
| 2016 | McNally [29] | Wavelet, RNN, LSTM | Predict the volatility of the cryptocurrency market with encoded feature by wavelet. |
| 2016 | Zbikowski [30] | EMA, SVM | Apply SL-based trading agent to cryptocurrency market |
| 2016 | Amjad [31] | EC, LDA | Propose a cryptocurrency market volatility prediction framework |
| 2017 | Jiang [32] | DQN | Apply deep RL-based trading agent to cryptocurrency market |

## 3. Proposed Method

In this section, we define the portfolio management problem formally and describe the detailed specification of proposed RL-based trading system that considers both profit maximization and risk restraint on portfolio management problem. The overall architecture of the proposed trading agent system is shown in Fig 2.

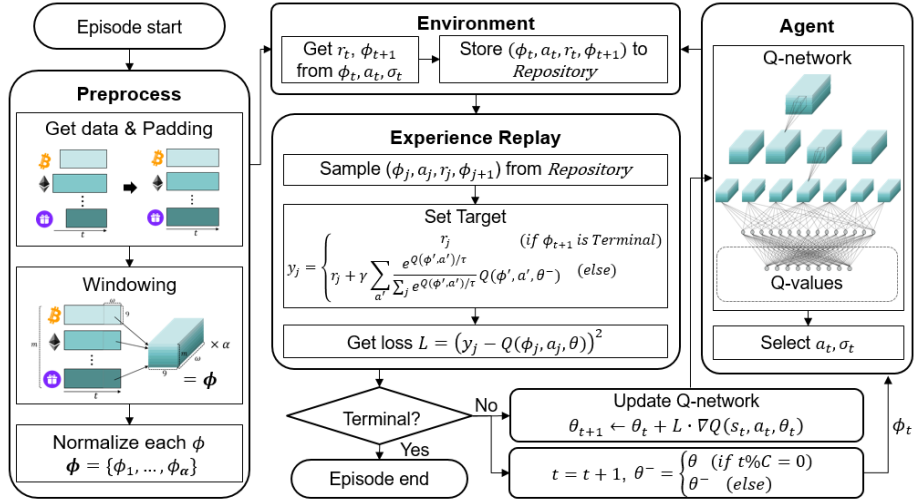

**Fig. 2. The architecture of the proposed trading agent system**

### 3.1. Problem Definition

The portfolio consists of $m$ assets and one base currency. In this study, we use the US dollar, called the US dollar tether (USDT) in the cryptocurrency market, as the base currency. Each asset has a price information vector $\mathbf{v}$ consisting of the opening price, the high price, the low price, and the closing price. Equation (1) represent the price information vector $v$ of the $m$-th asset.

$$\mathbf{v}_m = \{open\ price, high\ price, low\ price, close\ price\} \qquad (1)$$

The *price* $p_m$ of the $m$-th asset is defined as the average of the elements in the vector $\mathbf{v}$. Equation (2) is an expression for $p_m$.

$$p_m = \frac{1}{4}\sum_{i=1}^{4} v_{m,i} \qquad (2)$$

Where, more precisely, $p_m$ is the price of the base currency required to buy an asset of one base unit. The price vector $\mathbf{p}$ is defined as a vector that stores the price $p$ of all the assets. Equation (3) represents the form of $\mathbf{p}$.

$$\mathbf{p} = \{1, p_1, p_2, \ldots, p_m\} \qquad (3)$$

Where, the first element of $\mathbf{p}$ means the price for the base currency. The *portfolio vector* $w$ is a vector representing the amount of each asset. Equation (4) is an expression for $\mathbf{w}$.

$$\mathbf{w} = \{w_0, w_1, w_2, \ldots, w_m\} \qquad (4)$$

In a typical portfolio management problem, the initial portfolio vector $w_0$ holds only the base currency. Equation (5) is an expression for $\mathbf{w}_0$.

$$\mathbf{w}_0 = \{\delta, 0, 0, \ldots, 0\} \qquad (5)$$

where $\delta$ is the initial holding amount of the base currency. In the time step $t$, the total value $W_t$ of the portfolio is derived as the inner product of the price vector $\mathbf{p}_t$ and the portfolio vector $\mathbf{w}_t$. Equation (6) represents the $W_t$.

$$W_t = \mathbf{p}_t \cdot \mathbf{w}_t \qquad (6)$$

Finally, the job of the trading agent to solve the portfolio management problem is to maximize the profit $P_T$ at the terminal time step $T$. Equation (7) is the equation for profit $P_T$.

$$P_T = \frac{W_T}{W_0} \qquad (7)$$

### 3.2. Asset Data Preprocessing

In this work, we use the trade history of the assets of the cryptocurrency market(thereafter simply referred to as *coin*) as data. We use the API of the Binance Exchange to get the data [33]. The Binance

Exchange offers high security and has nearly 300 tradable coins, including bitcoin. In addition, despite its short history, among the nearly 10,000 cryptocurrency markets, the volume of day trading was the second highest [34].

Treating the trade history for all coins requires tremendous computational power. Therefore, before the preprocessing, $m$ coins with the highest trading volume are selected. This selection criterion is quite reasonable, since it means that the coin with a lot of daily trading volume is an active investment item.

Second, we collect data for the target period. The Binance API provides a vector containing nine trade history properties in every minute for each coin. Therefore, each coin has $(\alpha \times 9)$ size trade history matrix. Where $\alpha$ is the size of the target period converted to minutes. All elements of the trade history vector are shown in Table 2.

**Table 2. Property of trade history vector $\rho$**

| Element | Description |
| --- | --- |
| Open | Indicates the open price |
| High | Indicates the high price |
| Low | Indicates the low price |
| Close | Indicates the close price |
| Volume | Shows the trading volume. |
| Number of trades | Shows the number of trades |
| Asset volume 1 | Indicates the quote asset volume |
| Asset volume 2 | shows how much taker bought the base asset volume |
| Asset volume 3 | shows how much taker bought the quote asset volume |

Third, pads all other coins to match the matrix size of the longest coin. This is because the days when each coin was listed on the Binance Exchange are different. We use the zero-padding method. The theoretical basis of this method may be somewhat lacking, we have empirically confirmed that this method covers well the missing data.

Fourth, we stack all trade history matrices orthogonally to create one large block. Then, using the sliding window method with window size $\omega$, create $(\alpha - \omega + 1)$ 3D sequential blocks with $(\omega \times m \times 9)$ shape.

Finally, normalize each block using the min-max normalization method. These blocks are called *history block* $\phi$, and each history block is used as input(ie, state) of RL-based trading agent. Fig. 3 shows an example of visualizing the created history block.

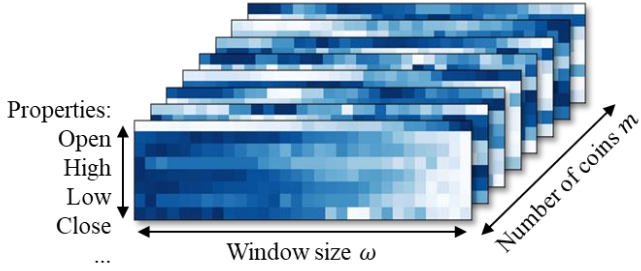

**Fig. 3. History block visualization**

### 3.3. RL-based trading system

When the portfolio management problem is expressed in the RL framework, the agent is considered as a portfolio manager that provides the trading strategy (ie, action) $a$ of the assets according to the current state $\phi$ of the capital market environment. The environment is a component that has all the trade histories for the assets. Also, here is where the assets in the portfolio are substantially traded, following the trading strategy provided by the agent. As an evaluation of the trading strategy, the environment returns the reward $r$ and provides the next state $\phi'$ to the agent again.

The proposed RL-based trading system follows the DQN structure [19]. Therefore, our agent obtains the q-values of each action through q-network (see section 3.3.3 for details), and environment stores an experience set $(\phi, a, r, \phi')$ in the repository. Then, update the q-network by batch-sampling the experience sequence randomly from the repository in the experience replay module.

#### 3.3.1 Action and Reward

Our agent determines not only the action $a$ for a given state, but also the ratio $\sigma$ at which the action is applied. First, the agent's action set is divided into three types: hold, buy, and sell. The buy and sell type actions are defined discretely for each asset. That is, an agent has (2m + 1) actions. Equation (8) represents action set A.

$$A \ni \begin{cases} none \\ buy\ w_b \\ sell\ w_b \end{cases}, where\ b \in \{1, \dots, m\} \qquad (8)$$

Where $w_b$ is an element of the portfolio vector defined in (4), and $none$ means that the portfolio vector values are maintained without any trading. Agent obtains q-values of each action through q-network and selects action by using $e$-greedy algorithm as behavior policy. Second, the ratio $\sigma$ to is defined as the softmax value for the q-value of each action. It means that the domain of $\sigma$ is a real number greater than or equal to 0 and less than or equal to 1. Therefore, the ratio at which the selected action is applied can be determined continuously. For example, if a strategy of buying a  $b$-th asset at a ratio of 0.5 is chosen, the  $b$-th asset is bought using 50% of the base currency. Equation (9) represents the ratio  $\sigma$.

$$\sigma = \frac{\exp(q_a)}{\sum_{a'} \exp(q_{a'})} \qquad (9)$$

The environment applies the trading strategy returned by the agent to the portfolio and returns the reward $r$. To obtain the reward, we define the variable ratio  $\eta = \beta(W_t/W_{t-1} - 1)$  for the total value of the portfolio before and after using the trading strategy at time step $t$. Where  $\beta$  is a variable ratio amplification constant that takes a positive integer as a domain. Also, the value of reward is clipped to [-1,1] to avoid overfitting. In conclusion, equation (10) represents reward  $r_t$  at time step $t$.

$$r_t = \begin{cases} 1 & if(\eta > 1) \\ -1 & if(\eta < -1) \\ \eta & else \end{cases} \qquad (10)$$

### 3.3.2 Target Policy

The target, which has the same role as the class in supervised learning, consists of reward and q-value for the target policy. The RL system based on the DQN structure traditionally uses the target based on the q-learning algorithm [35]. But, the greedy policy, which is the target policy of q-learning, considers only the optimal q-value for the next state. The process of greedy policy has the problem of ignoring the risks that may arise from exploration [36]. This problem is even more fatal for domains where it is important to choose safe actions, such as capital markets. Our the new target leads the agent to choose a safe action. The proposed target policy is a key contribution of this work.

The proposed system uses the target of the expected sarsa algorithm as the base frame [37]. This algorithm reflects the expected value of the q-value that can be obtained when following the target policy. It can also cover greedy policy [38]. Equation (11) represents the target of the expexted sarsa algorithm with function approximation.

$$Target_{E\_SARSA} = r + \sum_{a'} \pi(a'|s')Q(s',a',\theta^-) \qquad (11)$$

Where,  $\pi$  represents the target policy. And we use softmax algorithm with temperature term as target policy. This target policy adjusts the greediness according to the temperature  $\tau$, which has a positive real number as a domain. Equation (12) represents the target to which the proposed target policy is applied.

$$Target = r + \sum_{a'} \frac{e^{Q(s',a',\theta^-)/\tau}}{\sum_j e^{Q(s',a_j,\theta^-)/\tau}} Q(s',a',\theta^-) \qquad (12)$$

Equation (12) considers all other q-values as well as the optimal q-value that the agent can obtain in the next state. Also, when $\tau$ converges to 0, it is the same as the target using greedy policy. Thus, if $\tau$ is moderately small, the target is similar to the greedy policy, but a non-greedy q-values(ie, risk) are slightly taken into account. And this method satisfies the constraint of DQN which must use off-policy algorithm.

However, since the distribution of q-values is different for each state the greediness of each state can be drastically changed. For example, the softmax value for the optimal q-value between the state where the magnitude of the q-values is different by 10 and the state where the magnitude of the q-values is different by 0.01 is very different. This causes the learning to become unstable. Thus, to get similar greediness in all states, we redefine the temperature as the mean of the absolute values for all q-values in each state multiplied by hyper temperature. where hyper temperature $\tau'$ is the parameter determined by the user to determine the greediness. The redefined temperature rounds the difference in greediness between states. Equation (13) represents temperature $\tau$.

$$\tau = \frac{\sum_{i=1}^{m}|Q(s', a_i, \theta^-)|}{m} \times \tau' \tag{13}$$

### 3.3.3 Q-Network Structure

The q-network is a deep neural network(DNN) that takes a history block as input and returns the q-value of each action as output. In this work, we construct a q-network using the Convolutional Neural Network (CNN), which is a DNN method that hierarchically extracts local features through a weighted filter [39]. Our q-network consists of 3 convolutional layers (Conv) and 2 fully connected layers (Fc) and does not use pooling to keep the history block information. And, we convolve the history block using 3-dimensional filters. The detailed structure of q-network is shown in Table 3. Where, *m* is the number of assets.

Table 3. Q-network Structure

| Layer | Filter size | Stride | Number of filters | Activation |
|---|---|---|---|---|
| Conv1 | 6x2x3 | (1,1,1) | 32 | ReLu |
| Conv2 | 5x4x4 | (2,1,1) | 64 | ReLu |
| Conv3 | 3x3x3 | (2,1,1) | 64 | ReLu |
| Fc1 | - | - | 512 | ReLu |
| Fc2 | - | - | 2m+1 | Sigmoid |

## 4. Experimental Results

In this section, we describe how to experiment settings and tuning key hyper-parameters. After that, we verify performance through various experiments.

### 4.1 Experiment Setting

We selected 8 coins, including bitcoin, with high trading volume based on the data collection day. And through the Binance API, we collected data for the coins from August 2017 to March 2018. The data collection period is the period in which the price volatility of the coins fluctuates extensively as interest about the cryptocurrency market increases drastically. This implies that our experimental environment will be very risky for investment. We conduct a back-test with collected data to check the performance of the proposed method.

Three evaluation standards are used for the performance evaluation of the trading agent. The first is profit $P_T$ introduced in (7). This value represents the rate of return of the final profit based on the investment funds. Second, the Sharpe ratio (SR) is used to take the risk of an investment into account [40]. If the ratio is higher, it means that the portfolio management strategy is less risky. Equation (14) is the definition of the Sharpe ratio $S$.

$$S = \frac{P_T - P_F}{\rho_P} \quad (14)$$

Where, $\rho_P$ is the standard deviation of the expected return and $P_F$ is the return of a risk-free asset. In our case $P_F$ is 0. Because, risk-free assets are USDT and quoted currencies can also be converted into USDT. Third, the maximum drawdown (MDD) represents the maximum loss from a peak to a trough [41]. the lower the MDD, the more likely it is to provide a portfolio management strategy that can take the downward risk into account. Equation (15) represents the MDD mathematically.

$$MDD = \max_{\substack{\tau > t \\ t}} \frac{P_t - P_\tau}{P_t} \quad (15)$$

### 4.2 Hyperparameter Optimization

The proposed trading agent has a number of hyper parameters that need to be adjusted. The set of hyper parameters includes many hyper parameters used in the DQN structure, the window size for generating the history block, and the hyper temperatures introduced in (13). Among these, hyper temperature is a key determinant of the performance of the proposed system. Also, the window size has a large impact on decision of the trading agent. Because it determines the size of the market that can be considered at one time. Therefore, we conduct a traditional grid search to examine the tendency between two important hyper parameters for profit. The grid search experiment is performed during the test period from 2018/01/24 through 2018/02/25. In this period, the period in which the market price fluctuates sharply and the period in which the market is transversal have similar shares. So that, various situations of the market can be learned well. We averaged the profits from 10 back-tests per hyper parameter pair. Fig. 4 shows the results of a grid search experiment for two hyper parameters.

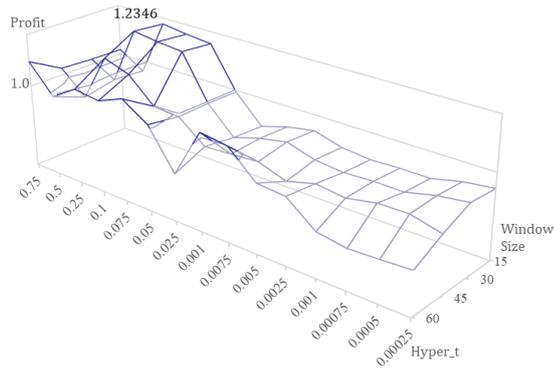

**Fig. 4. The grid search to find the optimal value for the two most important hyper parameters**

As a result, the trading agent obtained the highest profit at the point of window size of 30 and hyper temperature of 0.25. In addition, the overall experimental results showed a convex shape with the point as the apex. In all subsequent experiments, the above values are fixed. All other hyper parameters were determined through heuristics. The specifications of all hyper parameters are summarized in Appendix A.

### 4.3 Performance Comparison

**Table 4. Performance comparison of various portfolio management methods**

| Layer | SR | MDD | Profit |
|---|---|---|---|
| UBAH | 0.0132 | 0.6332 | 5.1587 |
| UCRP | 0.0156 | 0.4277 | 6.2377 |
| EG | **0.0207** | 0.4401 | 1.7552 |
| PAMR | 0.0138 | 0.4789 | 9.7058 |
| DQN(basic) | 0.0132 | 0.4321 | 7.3628 |
| DQN(ours) | 0.0153 | **0.3860** | **18.7674** |

For comparison with existing studies, two of the most commonly used benchmark algorithms, two portfolio management algorithms, and the basic DQN-based trading agent we implemented is compared with our trading agent. The first benchmark Uniform Buy and Hold (UBAH) is a strategy to uniformly invest in all assets and hold a portfolio until the end. The second benchmark Uniform Constant Rebalanced Portfolio(BCRP) is a baseline strategy which will rebalance the portfolio uniformly every trading period [42]. And, the two portfolio algorithms are Exponential Gradient(EG) [43] and Passive

Aggressive Mean Reversion strategy(PAMR) [27]. The hyper parameters of the two portfolio algorithms were set using the values recommended in each paper. The basic DQN-based trading agent simply uses the greedy policy as the target policy. We conducted the back-test during test period from 2017/11/17 to 2017/12/26. This period is part of the period during which the cryptocurrency market was most active. The performance comparison results of the methods are shown in Table 4.

Interestingly, we found that even if an agent followed only benchmark strategies, it gains over five times the profit during the test period. However, these benchmark strategies have resulted in over 60% drawdown in the worst case. This shows that during the test period, the prices of coins generally increased, and at the same time, the volatility of prices fluctuated drastically. The first portfolio management algorithm EG obtained the highest sharpe ratio among all the experimental cases. However, profit was significantly lower than the benchmark strategy. PARM made the second highest profit. However, other evaluation standards show that PARM has gained profit unstably. The trading agents using basic DQN stayed on average for all evaluation standards. The trading agent using the proposed DQN has the highest profit and lowest MDD among the methods used in the experiment. We also recorded the highest sharpe ratio among the methods that earned more than 7 times. These test results show that our agent can get high returns considering the risk of capital market.

**4.4 Robustness Verification**

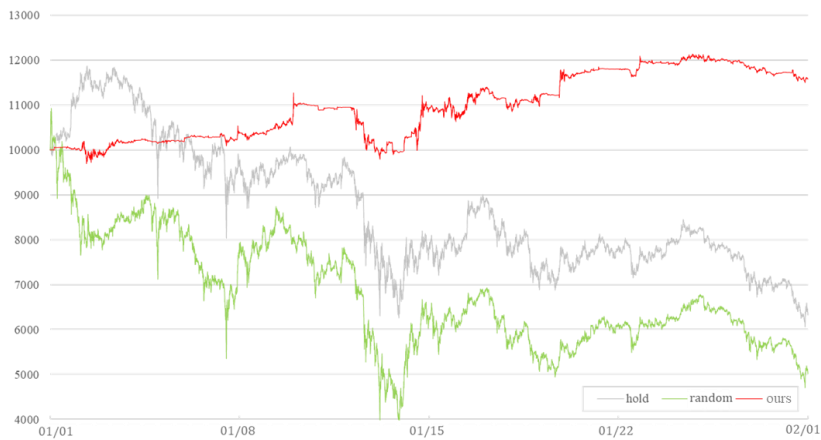

Fig. 5. Portfolio management results during the high-risk period

January 2018 is the period when the largest drawdown occurred in the history of the cryptocurrency market. We demonstrate that the proposed trading agent can firmly manage portfolio management against the risks in the capital market through back-testing experiments for this period. In this experiment, two baselines are used. The first baseline *hold* uses a strategy of holding up to the end of the assets it had in the beginning. The second baseline *random* randomly buys or sells assets without special strategies. A random strategy represents a risk that may arise when making investments without proper strategies.

Fig. 5 shows the experimental results of the two baselines and the proposed method during the test period. As a result, MDD of our trading agent reached only 13% while the MDD of hold reached 49% and the MDD of random reached 64%. And while the baselines lost their initial investment funds, our agents stably achieved a profit of about 1.15. Fig. 6 shows the difference between the hold and the average reward acquisition amount of the proposed trading agent. As shown in the figure, the proposed trading agent has a significantly narrower range of reward than hold. This indicates that our agent prefers actions that are not risky. This result also supports the claim that higher profits can be obtained when pursuing low risk [9].

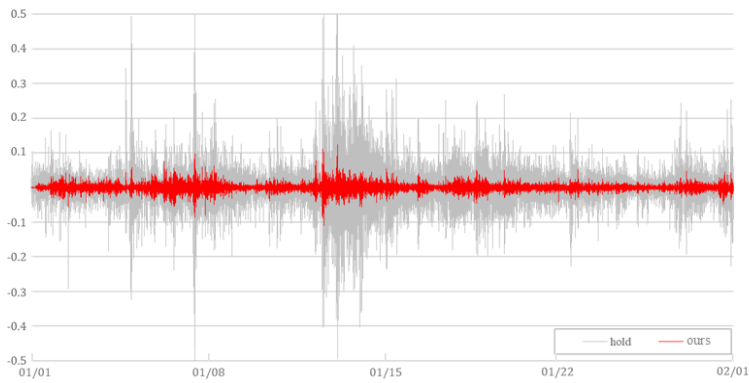

**Fig. 6. Average reward acquisition amount during the test period**

Second, we show that the proposed trading agent can generalize the learned information through experiments on unseen data. At the same time, we show that the agent is robust against the learning period while experimenting with setting the training period shorter and shorter. The first agent trains through data from 2017/08/01 to 2018/02/07. From the next agent, the training start time is delayed by one month. A total of five agents are tested, and the last agent trains through data from 2017/12/01 to 2018/02/07. All agents learn for 100 epochs, and the back-testing period is the same from 2018/02/07 to 2018/03/08. We also use the hold strategy as the baseline.

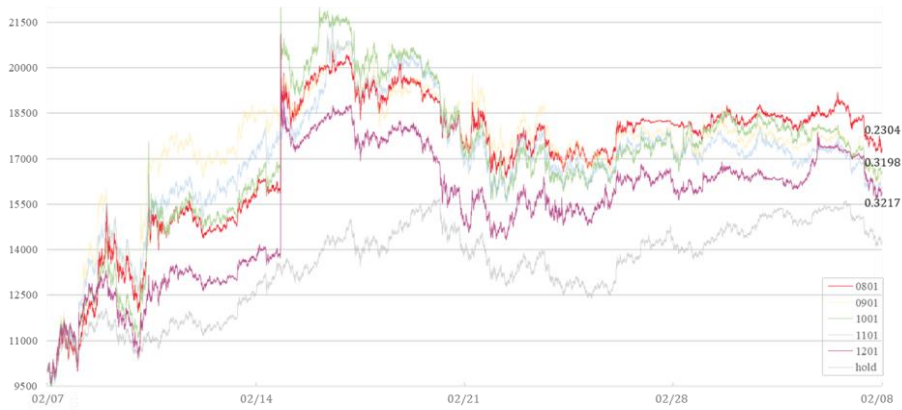

**Fig. 7. Results of robustness test during test period**

Fig. 7 shows the results of the experiment described above. As can be seen, all agents have a higher profit than hold for the unseen test period. This indicates that the generalization ability of the proposed trading agent is robust against the data size. However, the drawdown of the agent tended to increase as the training period became shorter. For example, the MDD of agents learned from December 1 reached 32%, while the MDD of agents learned from August 1 reached 23%.

## 5. Conclusions

This work proposes a deep RL-based trading agent that manages portfolios in the capital market, taking both profit maximization and risk restraint into account. And, this work proposed a new target policy, in order to allow the agent to learn that the action with the least risk is preferred. We conducted a back-test through cryptocurrency market data to confirm the performance of the proposed trading agent. As the result of the experiment, the proposed agent obtained the highest profit among the compared algorithms and provided the most stable investment strategy among the high profitable trading agents. In addition, another experiment showed that the agent can maintain robust generalized performance even if market volatility is large or training period is short.

However, our encoding method lacked the theoretical basis to successfully encode the information in the capital market. Therefore, we will study how to properly encode capital market information as a future work. Second, to improve the performance of the trading agent, we will study how to construct an improved q-network structure by combining various DNN methods.

# Appendix A. Hyper Parameters

### Table 5. Specification of all hyperparameters

| Name | Value | Description |
|---|---|---|
| Number of assets | 8 | Number of assets used for trade. |
| Window size | 30 | Size used for sliding window method |
| Memory size | 100000 | The capacity of the repository |
| Discount factor | 0.99 | Size of discount factor gamma used for q-network update |
| Minibatch size | 32 | Number of states learned per update |
| Update frequency | 4 | Number of actions selected by agent between update |
| Initial exploration | 1 | The initial $\varepsilon$ value of the $\varepsilon$-greedy policy |
| Final exploration | 0.1 | The last $\varepsilon$ value of the $\varepsilon$-greedy policy |
| Exploration annealing length | 1000000 | The step size required to reduce the initial $\varepsilon$ value to the final $\varepsilon$ value |
| Update start size | 20000 | The step size at which the q-network update begins |
| Target network update frequency | 20000 | The target Q-network update frequency ($C$) |
| Variable ratio amplification constant | 5 | Constants that amplify the portfolio variable ratio ($\beta$) |
| Hyper temperature | 0.25 | The value that determines the greediness of the target Q-network |